\documentclass[journal]{vgtc}                     

\onlineid{0}

\vgtccategory{Research}

\vgtcpapertype{please specify}

\title{Flow-Aware Diffusion for Real-Time VR Restoration: Enhancing Spatiotemporal Coherence and Efficiency}

\author{%
  \authororcid{Yitong \ Zhu}{0009-0007-5717-3390},
  \authororcid{Qianghong \ Dong},
  \authororcid{Guanxuan \ Jiang}{0009-0001-3686-6266}, \authororcid{Zhuowen \ Liang}{0009-0001-0217-1685} and \authororcid{Yuyang \ Wang*}{0000-0003-0242-8935}
}
\authorfooter{
  Yitong Zhu, Guanxuan Jiang, ZhuowenLiang, Yuyang Wang are with the Hong Kong University of Science and Technology (GuangZhou).
  Qianghong Dong is with the National University of Singapore.
  E-mail: yzhu162@connect.hkust-gz.edu.cn, e1350210@u.nus.edu , gjiang240@connect.hkust-gz.edu.cn, simonliang484@gmail.com, yuyangwang@hkust-gz.edu.cn
}

\abstract{%
  Cybersickness remains a critical barrier to the widespread adoption of Virtual Reality (VR), particularly in scenarios involving intense or artificial motion cues. Among the key contributors is excessive optical flow—perceived visual motion that, when unmatched by vestibular input, leads to sensory conflict and discomfort. While previous efforts have explored geometric or hardware-based mitigation strategies, such methods often rely on predefined scene structures, manual tuning, or intrusive equipment. In this work, we propose U-MAD, a lightweight, real-time, AI-based solution that suppresses perceptually disruptive optical flow directly at the image level. Unlike prior handcrafted approaches, this method learns to attenuate high-intensity motion patterns from rendered frames without requiring mesh-level editing or scene-specific adaptation. Designed as a plug-and-play module, U-MAD integrates seamlessly into existing VR pipelines and generalizes well to procedurally generated environments. The experiments show that U-MAD consistently reduces average optical flow and enhances temporal stability across diverse scenes. A user study further confirms that reducing visual motion leads to improved perceptual comfort and alleviated cybersickness symptoms. These findings demonstrate that perceptually guided modulation of optical flow provides an effective and scalable approach to creating more user-friendly immersive experiences. The code will be released at https://github.com/XXXXX (upon publication).
}

\keywords{Virtual Reality (VR) Video Restoration, Cybersickness, Optical Flow, Temporal Consistency, Mamba Architecture}

\graphicspath{{figs/}{figures/}{pictures/}{images/}{./}} 

\usepackage{tabu}                      
\usepackage{booktabs}                  
\usepackage{lipsum}                    
\usepackage{mwe}                       
\usepackage{amsmath}
\usepackage{amssymb}
\usepackage{float}

\usepackage{mathptmx}                  

\begin{document}


\firstsection{Introduction}

\maketitle
Virtual Reality (VR) technologies have been widely adopted across various domains, including education, healthcare, industrial design, and entertainment~\cite{kaminska2019virtual,10.1145/3680533.3697068,luo2024using}. However, one persistent challenge continues to hinder broader adoption: cybersickness. Commonly experienced as nausea, dizziness, or disorientation, this phenomenon remains a critical barrier to fully immersive experiences. Among its primary causes, the mismatch between visual and vestibular cues~\cite{Kim2020ADeepMotionSickness,keshavarz2015vection}, as described by the well-established sensory conflict theory~\cite{Nürnberger2021MismatchofVisual-Vestibular}, has been widely recognized. This mismatch is particularly pronounced in visually rich environments, where artificial motion cues dominate the user's field of view. In such scenarios, visual motion alone can evoke a strong illusion of self-motion, even in the absence of physical movement.

A major source of this sensory mismatch is unnatural or excessive visual motion, which can induce discomfort even when users are physically stationary. In particular, optical flow, which refers to the perceived motion of the visual scene relative to the observer~\cite{lefevre2009optical}, has been identified as a key visual trigger. High-intensity optical flow creates a compelling illusion of self-motion; when unaccompanied by corresponding vestibular input, it amplifies perceptual inconsistencies and leads to cybersickness. This effect is especially pronounced in peripheral regions of the visual field, where motion sensitivity is heightened. As such, modulating the structure and magnitude of optical flow in rendered content~\cite{sun2024pixel,kreinin2023highfidelity} has become a critical strategy for enhancing comfort across a wide range of immersive applications—not limited to virtual locomotion, but including diverse interactive and head-tracked experiences.

A variety of hardware and software strategies have been proposed to mitigate cybersickness, including vestibular stimulation and visual content adaptation. Hardware-based methods such as Galvanic Vestibular Stimulation (GVS)\cite{pradhan2022visual} and motion platforms\cite{langbehn2017application} aim to align physical and visual cues but are often intrusive and impractical for consumer use. Software approaches, including virtual visual anchors\cite{whittinghill2015nasum}, dynamic field-of-view restriction\cite{fernandes2016combating}, and peripheral blurring~\cite{budhiraja2017rotation}, modify rendered content to reduce perceived motion, though they can compromise immersion or visual quality. Notably, Lou et al. introduced a geometric simplification framework that reduces peripheral optical flow by manually editing high-motion regions of the virtual scene~\cite{lou2022general,lou2022geometric}. Their method segments scene geometry based on offline optical flow analysis and replaces complex elements with simplified shapes. While effective, it relies heavily on predefined meshes, manual tuning, and offline  and asynchronous preprocessing, limiting its scalability in dynamic or procedurally generated environments.

To overcome these limitations, we propose a lightweight, real-time, AI-based approach that automatically attenuates disruptive motion patterns at the image level—without requiring mesh editing, scene-specific adjustments, or additional hardware. Designed as a plug-and-play module, our solution integrates seamlessly into existing VR pipelines, enabling end-to-end cybersickness mitigation with low computational overhead and high scene compatibility. Unlike traditional handcrafted methods, our algorithm learns to identify and suppress perceptually salient optical flow features directly from rendered images. This data-driven strategy eliminates the need for scene-specific preprocessing, ensuring a broad generalization across procedurally generated or complex environments. Crucially, our approach preserves visual fidelity and user agency, avoiding common trade-offs such as degraded image quality or constrained interaction.

Beyond demonstrating consistent reductions in average optical flow, we conducted a user study to assess subjective effectiveness. Participants reported significantly improved visual comfort and decreased cybersickness symptoms, affirming that our system offers both measurable and experiential benefits. Together, these results position our method as a scalable and perceptually aware solution for next-generation immersive systems. Our contributions can be summarized as follows:

\begin{itemize}
\item A generalizable, plug-and-play AI-driven approach is introduced for cybersickness mitigation, operating directly at the image level without the need for mesh simplification, scene-specific adaptation, or manual intervention.
\item The proposed approach effectively reduces perceived motion discomfort by suppressing high-intensity optical flow in rendered content, while preserving scene fidelity and user control across a broad range of immersive scenarios.
\item The approach is validated through a combination of objective metrics and user studies, demonstrating significant reductions in optical flow and improved user-reported comfort across diverse virtual reality environments.
\end{itemize}

\section{Related Work}

\subsection{Cybersickness and Visual Motion: A Sensory Conflict Perspective}
As VR technologies become increasingly prevalent in education, healthcare, and entertainment, one of the major barriers to mainstream adoption remains cybersickness, which is a form of motion-induced discomfort typically experienced as nausea, dizziness, or disorientation~\cite{Kim2020Multisensory,Regan2017The}. A widely accepted explanation for cybersickness is the \textit{sensory conflict theory}, which attributes user discomfort to mismatches between visual and vestibular inputs~\cite{Nürnberger2021MismatchofVisual-Vestibular}.  When users perceive motion visually, for instance through large field-of-view displays or scene movement, a lack of corresponding physical feedback can result in a conflict. This conflict often leads to perceptual instability.

In immersive VR environments, this conflict is often intensified by dynamic visual stimuli that produce strong optical flow, particularly in peripheral vision, where motion sensitivity is heightened~\cite{lefevre2009optical}. Excessive or inconsistent optical flow can induce a compelling illusion of self-motion, aggravating sensory mismatch, and triggering discomfort. Moreover, visual artifacts such as frame skipping, ghosting, and temporal discontinuity further disrupt motion continuity and exacerbate user disorientation~\cite{Regan2017The}.

Given the real-time and high-resolution demands of modern VR systems often require high frame rates~\cite{kemeny2020getting, jung2021floor}, so that cybersickness remains a critical challenge. Addressing it requires not only improving video quality but also ensuring perceptual coherence between the visual stimulus and the user’s motion expectations. As such, motion-consistent rendering and optical flow suppression have become essential goals in optimizing VR content.

\subsection{Cybersickness and optical flow}

Temporal consistency is essential for comfortable video perception, particularly in immersive VR environments where users are highly sensitive to motion discontinuities. One of the most critical visual cues contributing to motion perception and even can be extended to cybersickness is \textit{optical flow}, which encodes pixel-wise motion between consecutive frames~\cite{lefevre2009optical}. Excessive or inconsistent optical flow, especially in peripheral vision, can induce the illusion of self-motion and lead to a sensory conflict when not accompanied by corresponding vestibular input~\cite{Regan2017The}. This mismatch is a well-established cause of cybersickness~\cite{Nürnberger2021MismatchofVisual-Vestibular}, and managing optical flow has thus become a central concern in perceptual stability research for VR.

From a modeling perspective, optical flow has been widely used as a prior to improve temporal continuity in video restoration tasks~\cite{lu2022priors,cheng2017segflow}. These methods perform well in scenarios involving fine-grained motion; however, their reliance on pre-computed or external flow estimates limits their ability to adapt to varied and rapidly changing motion patterns dynamically. Moreover, traditional flow-based models typically treat motion information as an auxiliary signal~\cite{henter2020moglow, zand2023flow} rather than a central learning objective. This often leads to suboptimal integration between motion cues and content reconstruction. Their architectures are rarely end-to-end trainable, which further restricts flexibility in high-frame-rate, wide-field VR applications where temporal coherence is paramount.

Recent advances in generative modeling, particularly diffusion-based frameworks, have begun to explore the integration of motion cues into the generation process. For example, Control-A-Video~\cite{chen2024controlavideocontrollabletexttovideodiffusion} introduces flow-conditioned denoising to improve motion controllability and visual stability in text-to-video synthesis. However, most existing models still lack explicit motion supervision, making them ill-suited for scenarios where motion fidelity is crucial, such as VR where even subtle inconsistencies between expected and observed motion can disrupt vestibular-visual alignment and induce disorientation.

Motivated by these limitations, we argue that optical flow should not merely serve as a refinement mechanism but instead be embedded as a core supervision signal. In this work, the proposed diffusion-based restoration framework directly incorporates \textit{flow-guided supervision} into the training objective. This design encourages the model to internalize the temporal structure, producing transitions that are both perceptually smooth and physically consistent. By explicitly modeling motion continuity, our approach addresses both visual fidelity and perceptual stability in high-frame-rate VR environments.

\subsection{Explicit Geometry-Based Flow Reduction}

One direct strategy for reducing cybersickness is to explicitly suppress optical flow through geometric manipulation of the visual scene. Such methods modify the structure or appearance of virtual content to reduce perceived motion, particularly in peripheral regions of the field of view where motion sensitivity is greatest~\cite{lefevre2009optical}. In general, these approaches can be divided into two categories: \textit{scene simplification} and \textit{visual anchoring}.

Scene simplification methods attempt to lower motion saliency by reducing visual complexity in high-flow areas. Lou et al.~\cite{lou2022general} proposed a framework that segments a scene based on optical flow magnitude and applies geometric simplification to the high-motion regions, typically in the periphery. Their follow-up work~\cite{lou2022geometric} rendered original and simplified scenes in concentric fields of view, preserving fidelity while suppressing disruptive motion. Similar efforts include deforming peripheral scene geometry to align with user motion direction, thereby minimizing perceived motion~\cite{bruder2011tuning, so2001metric}. Visual anchoring techniques, on the contrary, aim to reduce the sensation of motion by introducing static visual references. A classic example is the insertion of a \textit{virtual nose} into the user’s field of view, which serves as a stable reference point and helps ground the user’s perception~\cite{whittinghill2015nasum}. Another strategy is peripheral blurring during rapid motion events, which limits visual flow without interfering with central vision~\cite{budhiraja2017rotation}.

While effective in reducing cybersickness, these techniques often rely on manual scene access, require handcrafted thresholds, and are tailored to specific environments. Their lack of scalability, generalization, and adaptability to procedurally generated or real-time scenes limits their practical deployment in modern lightweight VR systems.

\subsection{Learning-Based Flow-Aware Modeling}

In contrast to explicit manipulation of scene geometry, learning-based methods seek to model and regulate motion perception through data-driven optimization. These approaches aim to internalize motion continuity and generate temporally coherent content that minimizes perceptual conflicts without manual intervention. Early flow-aware learning models, such as DAIN~\cite{bao2019depth} and FLAVR~\cite{kalluri2023flavr}, incorporate pre-computed optical flow to warp adjacent frames, reducing artifacts like ghosting and flicker. However, these methods treat flow as a secondary alignment signal, limiting adaptability and robustness in immersive settings. Moreover, they lack end-to-end differentiability, which restricts their generalization to unpredictable motion patterns or scene dynamics common in VR.

Recent advancements explore deeper integration of motion priors into generative architectures. Transformer-based models like VRT~\cite{liang2022recurrent} have demonstrated the ability to capture long-range temporal dependencies, enhancing consistency across time. However, their high computational demands and lack of explicit perceptual supervision make them unsuitable for real-time, high-frame-rate applications such as VR. Diffusion-based generative models have recently emerged as a promising alternative, offering both quality and flexibility. Control-A-Video~\cite{chen2024controlavideocontrollabletexttovideodiffusion} conditions denoising flow features to produce controllable, temporally stable sequences. However, even these models often omit explicit motion supervision, making them vulnerable to subtle but perceptionally significant inconsistencies.

To address these challenges, we propose a flow-guided diffusion model that incorporates optical flow as the core supervision target. By embedding motion consistency directly into the training objective, the model is encouraged to respect temporal structure and produce transitions that align with both physical dynamics and perceptual expectations. This approach enables real-time, high-fidelity restoration suitable for VR systems, while simultaneously enhancing visual comfort and motion stability.

\section{Methodology}

\begin{figure*}
    \centering
    \includegraphics[width=\linewidth]{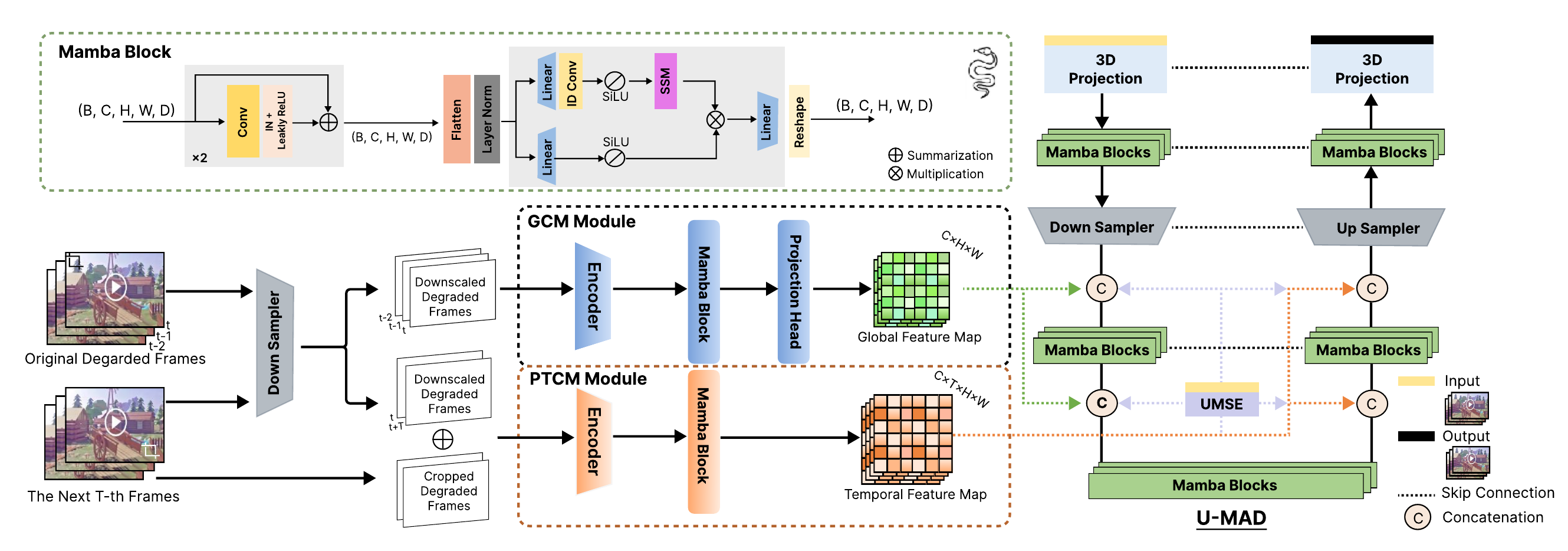}
    \caption{Framework of the proposed U-MAD}
    \label{U-MAD framework}
\end{figure*}

To more efficiently modulate the structure and magnitude of optical flow so as to reduce the cybersickness in real-time , we propose the \textbf{U-shaped Mamba Diffusion (U-MAD)} framework, which aims to restore high-quality video sequences from spatially and temporally degraded inputs. The framework, shown in Fig.~\ref{U-MAD framework}, takes as input a sequence of degraded frames $F_{\text{deg}} = \{f_1, f_2, \dots, f_T\}$ and a downsampled clean reference $F_{\text{raw}}$, available during training. The degraded frames are first spatially cropped into patches to improve processing efficiency, then passed through a U-shaped encoder–decoder backbone built on Mamba~\cite{gu2024mambalineartimesequencemodeling}, a state-space sequence model that enables scalable modeling of long-range temporal dependencies.

To supplement the local cropped input, a downsampled version of $F_{\text{raw}}$ is processed through a \textbf{Global Context Module (GCM)}. Simultaneously, a \textbf{Post-Temporal Context Module (PTCM)} captures temporal patterns by attending to future frames relative to each target frame. The two auxiliary modules, whose details are provided in Section~\ref{Sec:GTCM}, offer high-level spatial and temporal context to enhance the understanding of the cropped input sequence. To enforce motion consistency, we compute the optical flow between neighboring frames in $F_{\text{deg}}$ and encode it through a dedicated flow encoder. The resulting motion embeddings are injected into the Mamba backbone as conditional signals, guiding the denoising process toward temporally coherent and physically plausible reconstructions. Formally, the model predicts the restored output as:
\[
\hat{F}_{\text{res}} = \mathcal{D}_{\theta}(F_{\text{deg}}, F_{\text{raw}}, \text{Flow})
\]
where $\mathcal{D}_{\theta}$ represents the flow-conditioned diffusion denoising function implemented through the U-shaped Mamba architecture.

The final output $\hat{F}_{\text{res}}$ consists of a sequence of high-resolution, motion-consistent frames aligned with the ground truth target $F_{\text{raw}}$. Detailed descriptions of the core modules—including Mamba blocks, context pathways, and flow-guided conditioning—are presented in the following sections.

\subsection{Global and Temporal Context Modules}
\label{Sec:GTCM}

To mitigate the spatial and temporal limitations caused by cropped inputs and localized modeling, we introduce a dual-branch context encoding strategy consisting of a \textbf{Global Context Module (GCM)} and a \textbf{Post-Temporal Context Module (PTCM)}. These two modules provide complementary contextual information to enhance motion-aware and perceptually coherent video restoration.

The \textbf{GCM} addresses the mismatch between the full-resolution 360° stereoscopic VR video (e.g., 3840×2160) and the 512×512 cropped input patches used during training. Since each patch covers only a small portion of the scene, we introduce a spatial context pathway that encodes a downsampled version of the full degraded frame. This helps preserve global structural semantics and supplies auxiliary features to the main diffusion pipeline via conditional concatenation.

The \textbf{PTCM} is designed to enhance short-term temporal coherence by explicitly modeling the frame sequence following the target timestep. While the Mamba backbone captures long-range dependencies via a sliding window, flow-based control signals can be sensitive to local motion. The PTCM complements this by encoding temporal dynamics using only post-target frames (i.e., $t+1, t+2,...$), thereby improving short-range motion consistency without increasing memory overhead.

\begin{figure*}
    \centering
    \includegraphics[width=1.0\linewidth]{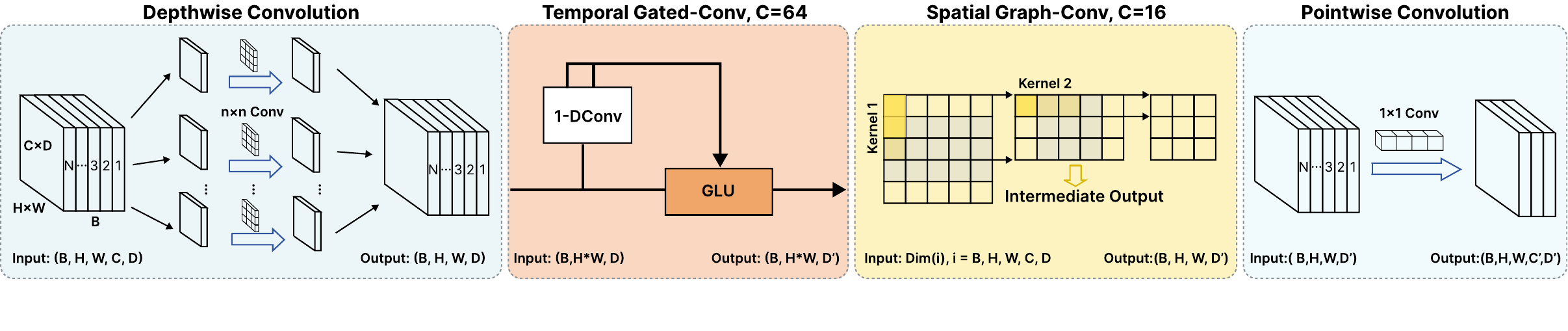}
    \caption{The detailed architecture of the TSC block.}
    \label{fig:tScblock}
\end{figure*}

Both GCM and PTCM are implemented using lightweight Temporal-Spatial Convolution (TSC) blocks. Each TSC block is built using depthwise separable convolutions to reduce computational cost, and the inputs to both modules are bicubically downsampled. To further improve efficiency and support deployment on mobile or low-power devices, we decouple temporal and spatial processing within the TSC design. As illustrated in Fig.~\ref{fig:tScblock}, each Temporal-Spatial Convolution (TSC) block is designed to separately process temporal and spatial dependencies, thereby reducing computational complexity while preserving motion-aware features. The input tensor has the shape $(B, H, W, C, D)$, where $B$ is the batch size, $(H, W)$ denote spatial resolution, $C$ is the number of channels, and $D$ is the temporal depth. We first reshape the tensor into $(B, H \times W, D)$ and apply a 1D convolution along the temporal axis to extract temporal dynamics. A Gated Linear Unit (GLU) is applied to enhance temporal selectivity. The result is projected via a pointwise (1×1) convolution back into a spatial format $(B, H, W, D')$. To model spatial dependencies, we perform a depthwise separable 2D convolution over $(H, W)$, followed by a pointwise convolution that restores the channel dimension. This decoupled architecture ensures minimal parameter overhead and supports real-time inference on low-power hardware. The final output has shape $(B, H, W, C', D')$, which can be aligned with the downstream backbone.

\subsection{U-shaped Mamba Diffusion Backbone}

The core of our architecture is a U-shaped diffusion network composed of hierarchical Mamba-based encoder and decoder blocks. Instead of traditional CNN or Transformer layers, we adopt \textbf{Mamba}~\cite{gu2024mambalineartimesequencemodeling}, a state-space sequence model that excels in modeling long-range temporal dependencies with linear computational complexity. Our implementation follows the recent advancements from Structured State Space Sequence Models (S4)~\cite{gu2022efficientlymodelinglongsequences} to the more selective and efficient u-Mamba (S6)~\cite{ma2024umambaenhancinglongrangedependency}.

Each Mamba block operates over spatio-temporal feature maps with shape $(B, C, H, W, D)$, where $B$ is the batch size, $C$ is the number of channels, $(H, W)$ are spatial dimensions, and $D$ is the temporal depth. The input features are normalized and transposed as necessary before being passed into the Mamba structure. Internally, each block includes two parallel branches designed to model complementary aspects of the input: The \textbf{first branch} captures global sequence features through pure linear transformation and activation, and the The \textbf{second branch} enhances local spatial awareness by applying convolution and a global-local attention mechanism (GLAM)~\cite{liu2021globalattentionmechanismretain}:
\begin{equation*}
    \text{Branch 1: } X_{1} = \text{SiLU}(\text{Linear}(\text{LayerNorm}(X)))
\end{equation*}
\begin{equation*}
    \text{Branch 2: } X_2 = \text{GLAM}(\text{SiLU}(\text{Conv}(\text{Linear}(\text{LayerNorm}(X)))))
\end{equation*}

The outputs from both branches are fused via element-wise multiplication, followed by a linear projection to restore the channel dimension:
\begin{equation*}
    X_{\text{out}} = \text{Linear}(X_1 \odot X_2)
\end{equation*}

This fusion strategy enables the model to combine global and local information effectively. The processed feature is then reshaped to its original 5D shape and propagated through the U-shaped structure.

In the encoder, stacked Mamba blocks progressively downsample the spatial resolution while increasing the temporal receptive field. Skip connections are preserved between symmetric layers in the encoder and decoder to facilitate hierarchical feature reuse. The Global Context Module (GCM) and Post-Temporal Context Module (PTCM) are injected at intermediate stages to enrich the encoder and decoder with auxiliary spatial and temporal information.

Finally, the denoising diffusion process is embedded within this U-shaped backbone. At each timestep $t$, a noisy representation $x_t$ is iteratively refined by the network toward the clean latent code $x_0$, under a temporally structured DDPM formulation~\cite{yang2024survey}. The detailed mechanism of conditional flow guidance is discussed in the following section.

\subsection{Unified Motion-Structure Embedding (UMSE)}

To enhance temporal consistency and motion plausibility during video generation, we introduce a \textbf{Unified Motion-Structure Embedding (UMSE)} module that integrates both \textit{optical flow cues} and \textit{location-aware structural priors} into the diffusion process. This embedding serves as a conditional input at every denoising timestep, guiding the network to generate temporally coherent and semantically consistent frames.

\paragraph{\textbf{Flow-guided Motion Encoding.}} 
UMSE extracts motion-aware features from pairs of consecutive degraded frames, denoted as $\{F_{t-2}, F_{t-1}\}$. Each frame is first passed through a lightweight convolutional encoder $g_\theta$ to obtain deep feature maps $g_\theta(F_t) \in \mathbb{R}^{H \times W \times C}$. A 4D pairwise correlation volume is then computed to measure pixel-level similarities:
\begin{equation}
    S(i,j) = \langle g_\theta(F_{t-2})_i, g_\theta(F_{t-1})_j \rangle,
\end{equation}
where $S \in \mathbb{R}^{H \times W \times H \times W}$ and $\langle \cdot, \cdot \rangle$ denotes the dot product. These correlations are used to iteratively refine an optical flow field $\mathbf{u}_t$ through a learned residual predictor:
\begin{equation}
    \mathbf{u}_t = \mathbf{u}_{t-1} + \text{update}(S, \mathbf{u}_{t-1}),
\end{equation}
where $\text{update}(\cdot)$ is a trainable function that estimates the residual flow based on the current similarity volume and prior flow.

\paragraph{\textbf{Location-aware Structural Encoding.}}
In addition to motion cues, UMSE incorporates structural priors inspired by LOST~\cite{dehaghi2024reversingdamageqpawaretransformerdiffusion}, including frame index, window scale, and encoding quality parameters. These factors are first normalized and embedded into a compact vector using dedicated MLPs, then spatially broadcasted to match the resolution of the flow features.

\paragraph{\textbf{Conditional Embedding Injection.}}
The motion and structural embeddings are concatenated and projected via a linear layer into a unified conditional representation, denoted as $\mathbf{e}_t^{\text{UMSE}} \in \mathbb{R}^{H \times W \times C'}$. Formally, this embedding is constructed as:
\[
\mathbf{e}_t^{\text{UMSE}} = \phi\big(\mathbf{u}_t, \psi(\mathbf{s}_t)\big),
\]
where $\mathbf{u}_t$ is the estimated optical flow at timestep $t$, $\mathbf{s}_t$ represents the location-aware structural priors (e.g., frame index, window scale, and encoding quality), $\psi(\cdot)$ is a multi-layer perceptron that embeds the structural priors, and $\phi(\cdot)$ is a fusion function implemented as a linear projection following concatenation.

At each stage of the U-shaped Mamba-based diffusion backbone, this unified embedding $\mathbf{e}_t^{\text{UMSE}}$ is conditionally injected into the Mamba blocks through concatenation within the attention pathway. During denoising, the network uses this conditional input to guide its prediction:
\[
\hat{x}_{t-1} = \epsilon_\theta(x_t, t, \mathbf{e}_t^{\text{UMSE}}),
\]
where $\epsilon_\theta$ denotes the denoising network. This conditioning mechanism enables the model to dynamically align the generative trajectory with the underlying motion patterns and scene structure, thus improving motion realism, temporal consistency, and overall perceptual stability.

\begin{figure}
    \centering
    \includegraphics[width=0.98\linewidth]{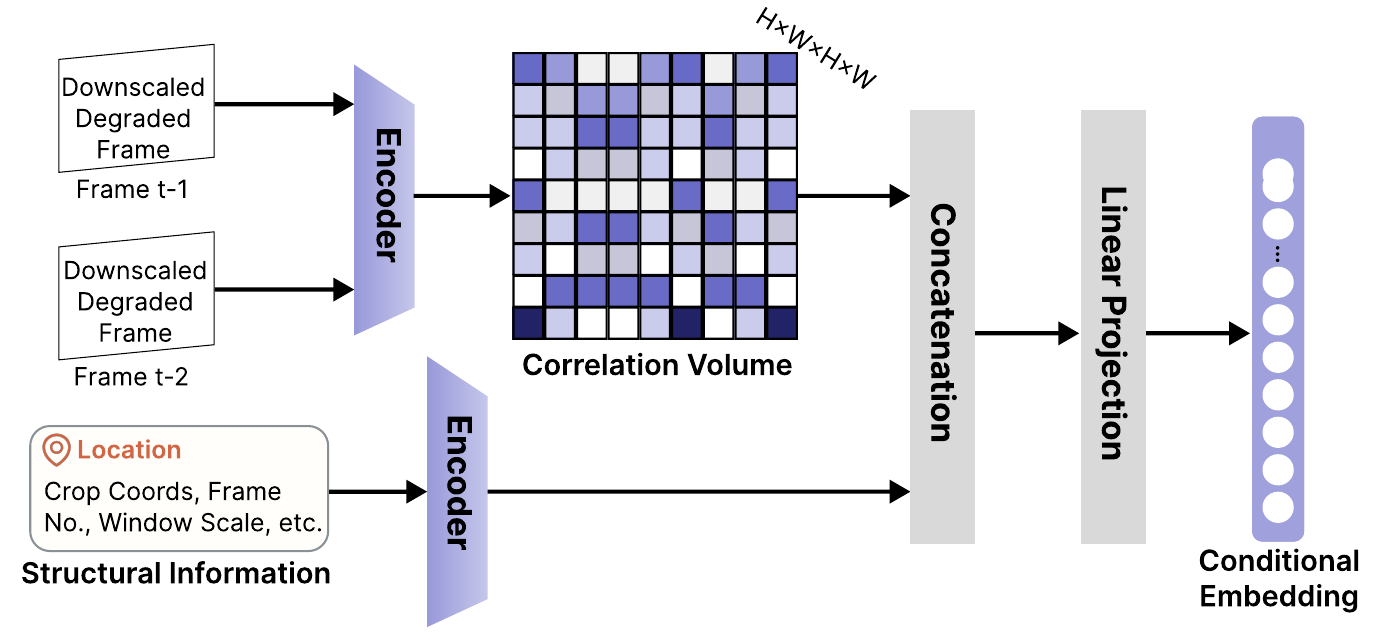}
    \caption{Overview of the UMSE module}
    \label{fig:architecture-of-umse}
\end{figure}

An overview of the proposed embedding pipeline is illustrated in Figure~\ref{fig:architecture-of-umse}, and we analyze the effectiveness of each component in Section~\ref{sec:ablation_studies}.

\subsection{Training Objective}

To jointly optimize spatial quality, temporal coherence, and motion consistency in video restoration, we define a composite loss function that supervises the network from multiple perspectives. The overall training objective integrates reconstruction, temporal alignment, and flow consistency constraints.

Let $\hat{F}_{\text{res}} = \{\hat{f}_1, \hat{f}_2, \dots, \hat{f}_T\}$ be the restored video output from the network, and $F_{\text{raw}} = \{f_1, f_2, \dots, f_T\}$ the corresponding clean reference sequence used during training. The total loss is defined as:
\begin{equation}
    \mathcal{L}_{\text{total}} = \lambda_1 \mathcal{L}_{\text{rec}} + \lambda_2 \mathcal{L}_{\text{temp}} + \lambda_3 \mathcal{L}_{\text{flow}}
\end{equation}

\begin{itemize}
    \item \textbf{Reconstruction Loss $\mathcal{L}_{\text{rec}}$}:  
    We use the Charbonnier loss~\cite{barron2019general} to robustly measure pixel-level differences between output and ground truth:
    \begin{equation}
        \mathcal{L}_{\text{rec}} = \frac{1}{T} \sum_{t=1}^{T} \rho(\hat{f}_t - f_t), \quad \text{where } \rho(x) = \sqrt{x^2 + \epsilon^2}
    \end{equation}

    where $\epsilon$ is a small constant (e.g., $10^{-3}$) for numerical stability. This loss emphasizes edge preservation and reduces sensitivity to outliers compared to L2.

    \item \textbf{Temporal Consistency Loss $\mathcal{L}_{\text{temp}}$}:  
    To encourage smooth transitions between adjacent frames, we apply a flow-guided warping loss:
    \begin{equation}
        \mathcal{L}_{\text{temp}} = \frac{1}{T-1} \sum_{t=1}^{T-1} \left\| \hat{f}_{t+1} - \mathcal{W}(\hat{f}_t, \hat{u}_t)
        \right\|_1
    \end{equation}
    where $\hat{u}_t$ denotes the estimated optical flow from $\hat{f}_t$ to $\hat{f}_{t+1}$, and $\mathcal{W}(\cdot, \cdot)$ is a differentiable warping function. This term penalizes temporal flickering and enforces frame-to-frame coherence.

    \item \textbf{Flow Consistency Loss $\mathcal{L}_{\text{flow}}$}:  
    We ensure the predicted motion is consistent with ground-truth or pre-computed flow. Let $u^{\text{gt}}_t$ be the reference flow (e.g., from RAFT), then:
    \begin{equation}
        \mathcal{L}_{\text{flow}} = \frac{1}{T-1} \sum_{t=1}^{T-1} \left\| \hat{u}_t -u^{\text{gt}}_t \right\|_1
    \end{equation}
    This constraint aligns motion dynamics in the generated video with physically plausible motion patterns.

\end{itemize}

We empirically set the loss weights as $\lambda_1 = 1.0$, $\lambda_2 = 0.5$, and $\lambda_3 = 0.1$ based on validation performance. Ablation studies in Section~\ref{sec:ablation_studies} evaluate the contribution of each component.

\section{Experiments}
\subsection{Datasets and Evaluation Metrics}
\subsubsection{Datasets}
The first dataset we use for training consists of 19 stereoscopic VR videos, as provided in~\cite{padmanaban2018TowardsaMachineLearningApproach}. The video sizes and lengths vary from one another. The width $\times$ height size of 9 videos is 2880 $\times$ 1080, while that of 8 videos is 1920 $\times$ 720, and the remaining two videos are 2730 $\times$ 1024. There are four videos of 61 seconds, four videos of 60.5 seconds, and 11 videos of 60 seconds. To test the robustness of our proposed model, we take another dataset of SEPE8K videos~\cite{Al2023SEPEDataset}. Both datasets' information is given in Table~\ref{tab:information-of-dataset}.
\begin{table}[h]
    \centering
    \caption{Overview of the Dataset Used for Training}\label{tab:information-of-dataset}
    \begin{tabular}{lccc}
        \toprule
        Dataset  & Dimension & Train set & Test set \\
        \midrule
        Padmanaban & 3D  & 15 & 4\\
        SEPE8K     & 2D     & 30 & 12  \\
        \bottomrule
    \end{tabular}
\end{table}

\subsubsection{Evaluation Metrics}
Our method uses a combination of the following metrics for quantitative evaluation:
\begin{itemize}
    \item \textbf{PSNR} and \textbf{SSIM}: widely used measures of pixel-level fidelity and structural similarity.
    \item \textbf{tOF} (temporal Optical Flow consistency): a temporal coherence metric that compares flow consistency between consecutive restored frames.
\end{itemize}

All metrics are averaged over the test set. We also report runtime and throughput in Section~\ref{sec:runtime} to evaluate efficiency under real-time constraints.

\subsection{Implementation Details}

Our model is implemented in PyTorch and trained end-to-end on six NVIDIA 4090 GPUs. The Adam optimizer with $\beta_1 = 0.9$, $\beta_2 = 0.999$, and an initial learning rate of $1 \times 10^{-4}$ is used in the training phase, which is decayed using a cosine annealing schedule over 200 epochs. The batch size is set to 8. All video frames are center-cropped to $512 \times 512$ resolution and sampled at 120 FPS. Each training sample consists of 7 consecutive degraded frames. We apply gradient clipping with a maximum norm of 1.0 to stabilize the training process. Mixed precision training is enabled via PyTorch AMP to reduce memory usage and accelerate convergence.

The U-shaped Mamba backbone comprises 4 downsampling and 4 upsampling stages, each with stacked Mamba blocks and residual skip connections. The Global Context Module (GCM) and Post-Temporal Context Module (PTCM) are implemented using lightweight Temporal-Spatial Convolution (TSC) blocks. Motion and structural priors are jointly encoded through our Unified Motion-Structure Embedding (UMSE) module. The motion encoder computes 4D correlation volumes between adjacent degraded frames and refines an optical flow field $\mathbf{u}_t$ via a trainable residual update function. Structural factors—including frame index, spatial window scale, and encoding quality—are embedded using dedicated MLPs and fused with motion features to form the conditional embedding injected at each Mamba layer.

The model uses a 25-step denoising diffusion process with a cosine noise schedule during both training and inference. On six 4090 GPUs, training takes approximately 12 minutes per epoch, and the model achieves a runtime speed of 115 FPS at $512 \times 512$ resolution.

\subsection{Comparison with Baseline Methods}

\subsubsection{Backbone Architecture Comparison}
\label{sec:backbonearchitecturecomparison}
We begin our experimental analysis by evaluating the impact of backbone architecture on the performance of diffusion-based video restoration. In particular, we compare three network designs used to implement the denoising process within the diffusion framework:
\begin{itemize}
    \item U-Net Diffusion: a 3D convolutional U-Net as proposed in VideoDiffusion~\cite{ho2022videodiffusionmodels}
    \item Transformer Diffusion: a DiT-style diffusion model adapted to video frames~\cite{peebles2023scalablediffusionmodelstransformers}
    \item Ours (U-MAD): our proposed U-shaped Mamba Diffusion backbone with motion-aware conditioning.
\end{itemize}

Quantitative results on the 360-VR test set are shown in Table~\ref{tab:backbone-comparison}. The U-MAD backbone significantly outperforms both CNN- and transformer-based counterparts across all evaluation metrics. It achieves a PSNR of 32.1, SSIM of 0.910, and the lowest LPIPS (0.113), indicating superior restoration quality and perceptual realism. Importantly, U-MAD also achieves the lowest temporal optical flow inconsistency (tOF), highlighting its strength in preserving smooth motion trajectories across frames.

\begin{figure*}[t]
\centering
\includegraphics[width=\textwidth]{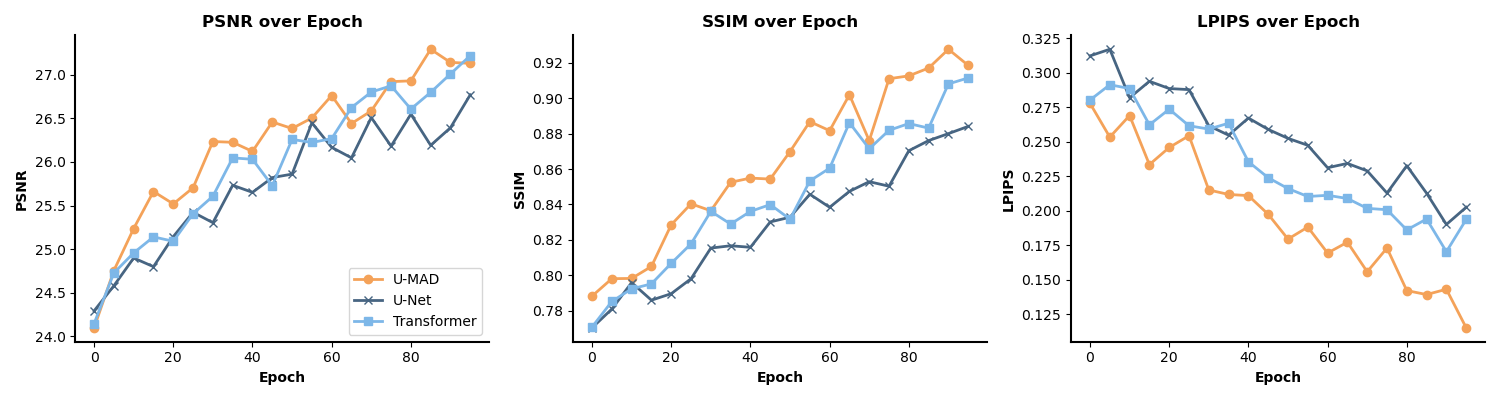}
\caption{
Performance over training epochs for each backbone. U-MAD exhibits more stable convergence and higher final accuracy across all metrics.
}
\label{fig:backbone-metrics}
\end{figure*}

Figure~\ref{fig:backbone-metrics} further shows the evolution of PSNR, SSIM, and LPIPS during training. U-MAD demonstrates faster convergence and reduced metric fluctuation, validating the benefits of our state-space backbone and flow-guided embedding in long-range temporal modeling.

\begin{table}[ht]
\centering
\caption{Backbone Comparison Results on 360-VR}
\label{tab:backbone-comparison}
\begin{tabular}{lcccc}
\toprule
\textbf{Method} & PSNR ↑ & SSIM ↑ & LPIPS ↓ & tOF ↓ \\
\midrule
U-Net~\cite{ho2022videodiffusionmodels} & 29.2 & 0.875 & 0.168 & 0.132 \\
Transformer~\cite{peebles2023scalablediffusionmodelstransformers} & 30.3 & 0.884 & 0.150 & 0.121 \\
\textbf{Ours (U-MAD)} & \textbf{32.1} & \textbf{0.910} & \textbf{0.113} & \textbf{0.095} \\
\bottomrule
\end{tabular}
\end{table}

\subsubsection{Conditioning Strategy Comparison}
Many recent studies have leveraged camera movement or other signals to guide video restoration. For instance, MotionCtrl~\cite{wang2024motionctrlunifiedflexiblemotion} and CameraCtrl~\cite{he2024cameractrlenablingcameracontrol} utilize camera motion as a control signal. However, one of the key factors contributing to cybersickness is the inconsistency in motion between frames. To address this, we incorporated optical flow consistency into our model, which helps maintain smoother motion and reduces the likelihood of inducing user discomfort.

\begin{figure*}[!h]
\centering
\includegraphics[width=\linewidth]{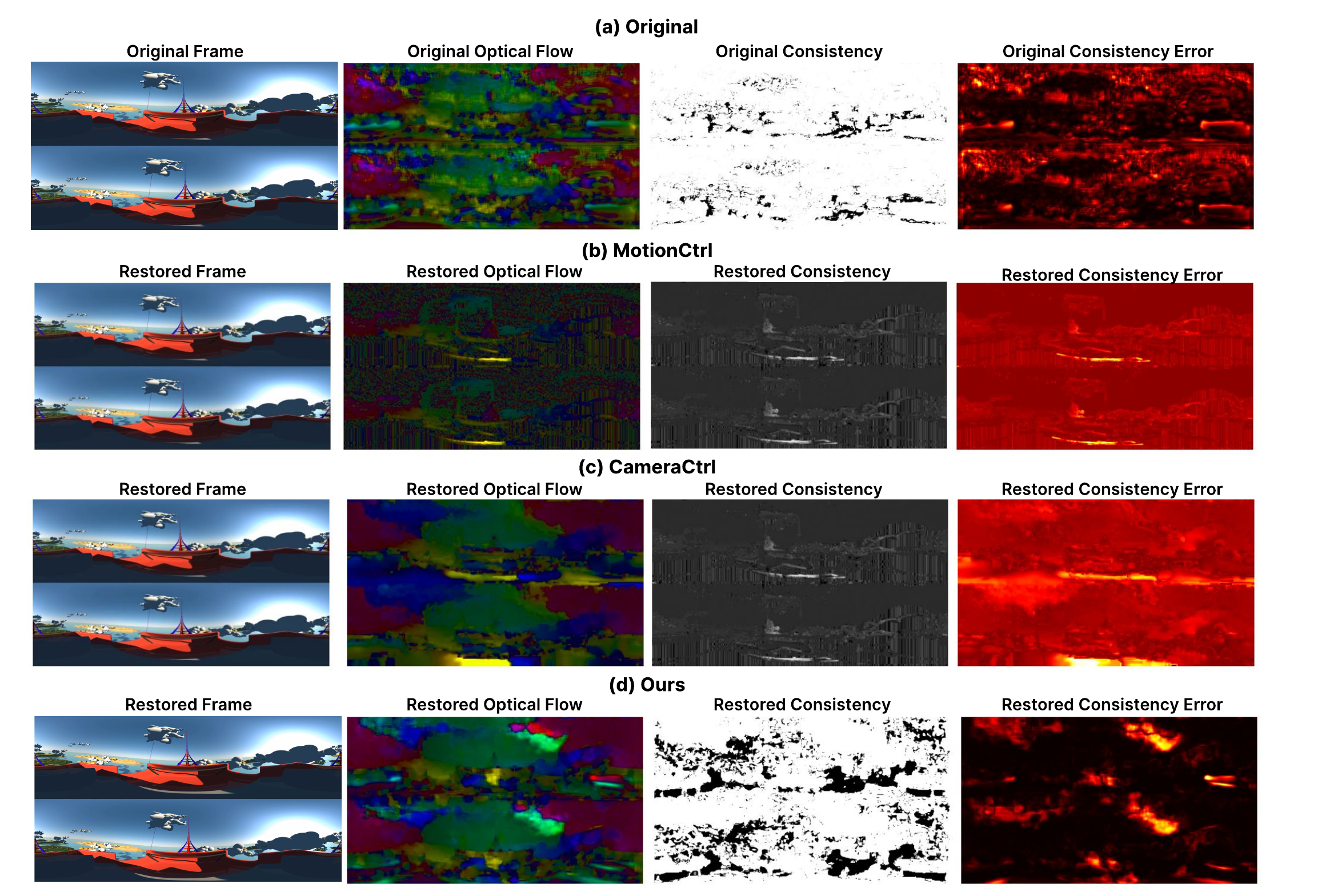}
\caption{
Visual comparison of different conditioning strategies: (a) Original degraded input; (b) Motion-based condition (MotionCtrl); (c) Camera parameter condition (CameraCtrl); (d) Our flow-guided condition. 
Each row shows restored frames, optical flow, temporal consistency, and consistency error maps. Our method achieves superior temporal alignment and smoother motion trajectories.
}
\label{fig:conditioning-visual}
\end{figure*}

To better understand the impact of different conditioning strategies, we present qualitative comparisons in Figure~\ref{fig:conditioning-visual}. The figure is structured in four columns, each representing one conditioning setup: (a) Original input (unrestored), (b) Motion-based conditioning, (c) Camera-based conditioning, and (d) Our proposed flow-based conditioning. Within each column, we visualize four components: the restored frame, the estimated optical flow, a computed frame-to-frame consistency map, and the corresponding consistency error map. The consistency map measures the similarity between adjacent frames, while the consistency error highlights areas of temporal flicker or misalignment. Notably, the error map in column (d) shows significantly fewer high-error regions (indicated in red) compared to (b) and (c), especially along object boundaries and in regions of large motion. This visual evidence supports our quantitative findings, indicating that pixel-level flow guidance provides better temporal stability and alignment under wide-field dynamic conditions.

\begin{table}[ht]
\centering
\caption{
Quantitative comparison of different conditioning strategies.
}
\label{tab:conditioning-comparison}
\begin{tabular}{lcccc}
\toprule
\textbf{Condition Type} & \text{PSNR $\uparrow$} & \text{SSIM ↑} & \text{LPIPS ↓} & \text{tOF ↓} \\
\midrule
Motion Vectors~\cite{wang2024motionctrlunifiedflexiblemotion} & 30.9 & 0.892 & 0.135 & 0.115 \\
Camera Params~\cite{he2024cameractrlenablingcameracontrol} & 30.1 & 0.886 & 0.142 & 0.123 \\
\textbf{Ours (Flow-Guided)} & \textbf{32.1} & \textbf{0.910} & \textbf{0.113} & \textbf{0.095} \\
\bottomrule
\end{tabular}
\end{table}

\subsection{Domain-specific Effectiveness and Generalization} 
To evaluate how well our method generalizes in different video domains, we conduct experiments on both conventional forward-view datasets and VR-specific omnidirectional datasets. These domains differ significantly in motion characteristics, field-of-view (FOV), and spatial continuity. Conventional 2D video datasets (e.g., SEPE8k~\cite{Al2023SEPEDataset}) contain standard perspective videos, whereas VR datasets, like Padmanaban~\cite{padmanaban2018TowardsaMachineLearningApproach}, feature wide-field scenes with strong camera and object motion. This setup allows us to examine both generalization and domain-specific effectiveness. Table~\ref{tab:domain-comparison} summarizes the performance of our model in both domains. While our method performs well in conventional videos, its advantage becomes more pronounced in VR settings, particularly in LPIPS and tOF. These results indicate that motion-aware conditioning and Mamba-based temporal modeling are especially beneficial in wide-field, high-motion environments typical of VR content.

\begin{table}[ht]
\centering
\caption{Performance Comparison on Conventional vs. VR Domains}
\label{tab:domain-comparison}
\begin{tabular}{lcccc}
\toprule
\textbf{Dataset} & PSNR ↑ & SSIM ↑ & LPIPS ↓ & tOF ↓ \\
\midrule
SEPE8k & 31.5 & 0.901 & 0.125 & 0.110 \\
Padmanaban             & \textbf{32.1} & \textbf{0.910} & \textbf{0.113} & \textbf{0.095} \\
\bottomrule
\end{tabular}
\end{table}
These findings validate the domain robustness of our method while also highlighting its specialization in immersive video restoration. The enhanced performance in VR environments demonstrates the effectiveness of the proposed UMSE module in handling complex motion and wide-FOV challenges, which are less prominent in conventional datasets. This supports the broader applicability of U-MAD in real-world VR content delivery systems.

\subsection{Runtime and Efficiency Analysis} \label{sec:runtime}
We first evaluate the inference speed of the model in terms of frames per second (FPS) and per-frame latency. All models are benchmarked on a single NVIDIA 4090 GPU with a batch size of four and an input resolution of $512 \times 512$. Our U-MAD achieves a runtime speed of 115 FPS, significantly faster than Transformer and DiT-based backbones, which have already described in Sec~\ref{sec:backbonearchitecturecomparison}. In addition to speed, we compare the model size and parameter counts. U-MAD contains 32.8M parameters, whereas DiT-based diffusion models typically exceed 100M due to stacked self-attention layers. The comparison in Table~\ref{tab:runtime-comparison} shows that our lightweight state-space formulation enables comparable or superior performance with substantially fewer parameters, making it more suitable for deployment in VR headsets and edge devices.

\begin{table}[ht]
\centering
\caption{Runtime and Efficiency Comparison at 512×512 Resolution (NVIDIA RTX 4070)}
\label{tab:runtime-comparison}
\begin{tabular}{lccc}
\toprule
\textbf{Method} & FPS ↑ & Params (M) ↓ & GPU (GB) ↓ \\
\midrule
U-Net~\cite{ho2022videodiffusionmodels} & 45 & 132.7 & 7.2 \\
Transformer~\cite{peebles2023scalablediffusionmodelstransformers} & 23 & 220.3 & 11.6 \\
\textbf{Ours (U-MAD)} & \textbf{113} & \textbf{98.4} & \textbf{5.4} \\
\bottomrule
\end{tabular}
\end{table}

\subsection{Ablation Studies} \label{sec:ablation_studies}
\subsubsection{Effect of Key Modules}
We evaluate the contribution of each key component by progressively removing it from the full model. As shown in Table~\ref{tab:ablation-modules}, removing the motion-structure embedding (UMSE) causes a significant drop in temporal consistency (tOF ↑), while removing the context modules (GCM/PTCM) mainly degrades perceptual quality (LPIPS ↑). These results confirm that both motion-aware and structure-aware signals are crucial to high-quality restoration.
\begin{table}[h]
\centering
\caption{Ablation study on the effect of different modules; each row removes or replaces one component from the full U-MAD architecture.}
\label{tab:ablation-modules}
\begin{tabular}{lcccc}
\toprule
\textbf{Variant} & PSNR ↑ & SSIM ↑ & LPIPS ↓ & tOF ↓ \\
\midrule
Full U-MAD (Ours)               & \textbf{32.1} & \textbf{0.910} & \textbf{0.113} & \textbf{0.095} \\
– UMSE    & 30.6 & 0.894 & 0.131 & 0.124 \\
– GCM / PTCM       & 30.9 & 0.891 & 0.138 & 0.105 \\
– Flow injection only           & 31.2 & 0.898 & 0.125 & 0.111 \\
– All conditioning      & 30.2 & 0.884 & 0.142 & 0.130 \\
\bottomrule
\end{tabular}
\end{table}

\subsubsection{Condition Injection Strategy}
To evaluate the impact of different conditioning mechanisms, several strategies for injecting motion-structure embeddings into the diffusion process are compared:

\begin{itemize}
    \item \textbf{No Injection}: the baseline where no external condition (flow or layout) is used.
    \item \textbf{Late Fusion}: optical flow and layout embeddings are added at the decoder stage via attention gating.
    \item \textbf{Early Fusion (Ours)}: condition embeddings are injected at every layer of the Mamba backbone through concatenation, allowing fine-grained motion alignment from early stages.
\end{itemize}

Quantitative results are shown in Table~\ref{tab:ablation-injection}. Our early fusion strategy consistently outperforms others across PSNR, LPIPS, and tOF, demonstrating that tight integration of conditional priors and feature encoding is crucial for stable and perceptually coherent video restoration.

\begin{table}[ht]
\centering
\caption{Comparison of Different Condition Injection Strategies}
\label{tab:ablation-injection}
\begin{tabular}{lcccc}
\toprule
\textbf{Injection Strategy} & PSNR ↑ & SSIM ↑ & LPIPS ↓ & tOF ↓ \\
\midrule
No Injection & 30.6 & 0.891 & 0.135 & 0.117 \\
Late Fusion  & 31.2 & 0.898 & 0.127 & 0.110 \\
\textbf{Early Fusion (Ours)} & \textbf{32.1} & \textbf{0.910} & \textbf{0.113} & \textbf{0.095} \\
\bottomrule
\end{tabular}
\end{table}

\subsection{User Study}

To evaluate the effectiveness of our proposed video restoring technique in reducing cybersickness during VR experiences, we conducted a user study. This section aimed to determine whether restored videos, processed through U-MAD, could alleviate these symptoms compared to the original videos.

\subsubsection{Participant}
A total of 18 participants (7 male, 11 female; age range: 19-29 years, $M =  23.17$, $SD = 1.95$) were recruited through social medias. Inclusion criteria included normal or corrected-to-normal vision and no history of severe cybersickness or vestibular disorders. Fifty percent reported no prior VR experience. All participants provided informed consent after receiving complete information about the experimental procedures and objectives, including their right to withdraw at any time. Before the experiment, we obtained approval from the Ethics Department of our University with the protocol No. HKUST(GZ)-HSP-2024-0021.

\subsubsection{Setting}
To control for potential order effects, a counterbalanced design was employed. Participants were randomly assigned to one of two viewing orders: half (n = 9) watched the restored video first, followed by the original version; the other half (n = 9) viewed them in the reverse order.

Two video clips (original and processed) were selected from VR-compatible gaming environments, which is a segment of the roller coaster game, each lasting 2 minutes and rendered at a resolution of 4K (3840×2160) to match typical VR headset performance standards. Videos were played on an Oculus Quest Pro headset, calibrated individually for each participant to ensure optimal visual quality. The testing environment consisted of a quiet, dimly lit room with participants seated in adjustable chairs to minimize external distractions. 

Between different video viewing sessions, all participants were given a rest period of approximately one hour to ensure they had recovered from any symptoms of sickness caused by visual stimuli. Following each video viewing session, all participants were instructed to complete a standard Simulator Sickness Questionnaire (SSQ) \cite{kennedy1993simulator}.

\subsubsection{Result}

We conducted Paired Sample t-test and effect size for the metrics derived from the SSQ after verifying that the data satisfied assumptions of normal distribution (Shapiro-Wilk test, $p < .05$) and homogeneity of variance (Levene test, $p < .05$). 

\begin{figure}
    \centering
    \includegraphics[width=0.9\linewidth]{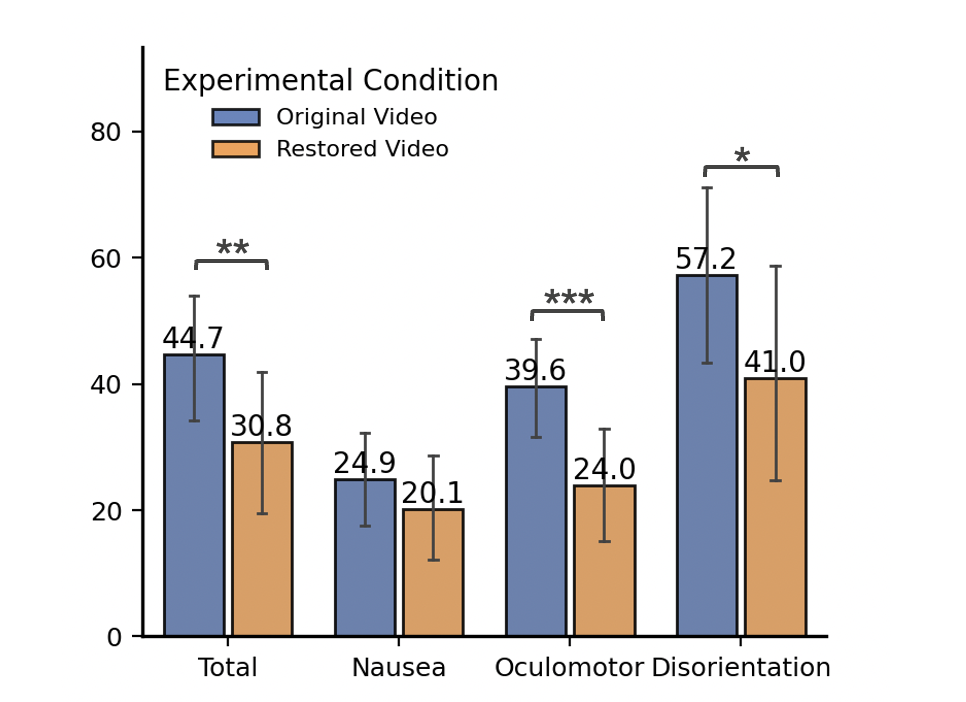}
    \caption{Comparison Between Original and Restored Videos}
    \label{fig:user}
\end{figure}

The within-subject analysis revealed that the processed videos demonstrated significantly lower risk of inducing cybersickness compared to the original videos. As illustrated in Fig.~\ref{fig:user}, the analysis of the results reveals that the participants reported significantly lower ($t=  3.7999, p < .01, Cohen's\text{ }d = .591$) overall SSQ socre after viewing the restored videos ($M = 30.75, SD =25.61$) than viewing the original videos ($M = 44.67, SD = 21.29$). The participants reported significantly lower oculomotor ($t= 5.1397, p < .01, Cohen's\text{ }d = .840$) after viewing the restored video ($M = 24.00$, $SD = 19.84$) than viewing the original videos ($M = 39.58$, $SD = 17.16$). The participants reported significantly lower ($t= 2.4658, p = .0246, Cohen's\text{ }d = .471$) disorientation  difficulty after viewing the restored videos ($M = 40.99, SD = 37.74$) than viewing the original videos ($M = 57.23, SD = 30.90$).

\section{Discussion}
This study set out to mitigate cybersickness in immersive VR environments by suppressing optical flow patterns that are disruptive to human perception. Our findings indicate that this goal has been successfully achieved. By incorporating a lightweight, plug-and-play AI module into standard VR pipelines, the proposed approach demonstrates that motion-aware video processing can effectively reduce sensory conflicts—without relying on mesh-level scene data, manual geometric editing, or specialized hardware. The central insight is that optical flow, when reframed as a perceptual control signal rather than a byproduct of scene geometry, provides a powerful means of enhancing temporal coherence and visual comfort in real-time immersive experiences.

At the algorithmic level, the proposed U-MAD model integrates a motion-aware architectural design with a diffusion-based generative framework, demonstrating substantial improvements over conventional baselines such as U-Net Diffusion~\cite{zhou2020unetredesigningskipconnections} and DIT~\cite{ho2022videodiffusionmodels}. This performance boost is primarily attributed to the incorporation of optical flow signals as explicit guidance during generation, which serve as strong priors for inter-frame motion estimation. Prior studies have emphasized the role of optical flow in maintaining motion continuity~\cite{koroglu2024onlyflowopticalflowbased}, and the results further validate this principle. In particular, U-MAD consistently improves temporal coherence and suppresses flicker and ghosting artifacts, especially in visually complex or fast-moving scenes. These improvements are evident in both quantitative metrics and user feedback. Participants in the user study reported noticeably reduced symptoms of dizziness, nausea, and disorientation, indicating that the motion consistency achieved by U-MAD is effective not just in algorithmic terms but also in enhancing real-world perceptual comfort. Compared to traditional approaches, such as field-of-view restriction~\cite{alfano1990restricting}, peripheral blurring~\cite{budhiraja2017rotation}, or scene simplification~\cite{han2021deep}, the proposed method preserves visual fidelity, interaction freedom, and scene detail, thereby avoiding the trade-offs commonly associated with discomfort-mitigation techniques. Moreover, it generalizes effectively across procedurally generated and dynamic environments, reinforcing its utility in real-world deployment.

While the proposed approach demonstrates clear advantages, it also faces several practical challenges that warrant further discussion. The training process, while conducted offline, incurs significant computational cost due to the combination of diffusion-based generation and dense optical flow estimation. This resource requirement aligns with known challenges in deploying deep generative models for edge AI~\cite{francy2024edgeaievaluationmodel, Loven2019EdgeAI}. To address this, we recommend a cloud-assisted pipeline wherein large-scale training is followed by compression techniques such as knowledge distillation, quantization, or structured pruning for efficient deployment on resource-constrained devices. Another practical consideration is inference latency, which is heavily influenced by the complexity of the optical flow estimator. While high-fidelity flow models enhance visual quality, they introduce substantial computational overhead, presenting a trade-off between performance and responsiveness. Additionally, optical flow alone may be insufficient for modeling complex scene dynamics such as object deformation, rapid camera motion, and occlusion. 

One potential direction for future research is to improve evaluation protocols that better capture perceptual responses to motion. While existing metrics are effective for assessing low-level image fidelity, they often fall short in measuring temporal continuity and stability—factors critical to user comfort in immersive experiences. To advance the field, we advocate for the development of temporally-aware and perception-driven benchmarks that align more closely with human sensitivity to motion artifacts, especially under real-time and interactive VR conditions. This approach may be further improved by incorporating additional motion cues beyond optical flow. Incorporating signals such as camera pose estimation, object tracking, and semantic motion segmentation may help the model better capture complex scene dynamics. Advances in multimodal fusion and cross-attention architectures offer promising pathways for combining visual, spatial, and semantic inputs into unified motion representations, thereby supporting broader generalization across diverse video domains.

U-MAD demonstrates a promising direction for perceptually guided video generation in immersive environments. Rather than forcing users to adapt to system constraints, this method adapts the visual content to human perceptual needs—flexibly, adaptively, and in real time. By bridging algorithmic generation with perceptual principles, this work opens new opportunities for effective and user-centered cybersickness mitigation.

\section{Conclusion}
This paper presents U-MAD, a motion-aware diffusion-based framework for mitigating cybersickness in immersive video applications through perceptually guided optical flow suppression. Departing from hardware-centric or geometry-dependent solutions, our approach operates entirely at the image level and generalizes across diverse environments without compromising visual fidelity or user agency. Through a combination of flow-guided supervision and plug-and-play architectural design, U-MAD achieves substantial improvements in both objective temporal consistency and subjective visual comfort. These results highlight the potential of aligning video generation models with perceptual principles to address long-standing challenges in VR systems.

Despite its advantages, U-MAD faces limitations in computational efficiency and expressiveness. Future work will explore lightweight, domain-adaptive optical flow networks to reduce inference overhead, as well as multi-modal motion cues such as camera pose, object dynamics, and semantic priors to enhance scene understanding. Additionally, we advocate for the development of temporally-aware, perceptually driven evaluation metrics to better align model assessment with human sensitivity to motion artifacts.
\bibliographystyle{abbrv-doi}
\bibliography{template}

\end{document}